# Bulk photovoltaic effect in two-dimensional ferroelectric semiconductor α-In$_2$Se$_3$

Chen xiaojuan[1], Xu kang[2], Qin Tingxiao[1], Wang Yubin[2], Chen Yuzhong[1], Liu Haiyun[1*], and Xiong Qihua[1,2*]

[1] Beijing Academy of Quantum Information Sciences, Beijing 100193, P.R. China;
[2] State Key Laboratory of Low-Dimensional Quantum Physics and Department of Physics, Tsinghua University, Beijing 100084, P.R. China
(*Electronic mail: chenxj@baqis.ac.cn)

**Abstract**

Bulk photovoltaic effect, which arises from crystal symmetry-driven charge carrier separation, is an intriguing physical phenomenon that has attracted extensive interest in photovoltaic application due to its junction-free photovoltaic and potential to surpass Shockley–Queisser limit. Whereas conventional ferroelectric materials mostly suffer from extremely low photocurrent density and weak photovoltaic response at visible light wavelengths. Emerging two-dimensional ferroelectric semiconductors with coupled visible light absorption and spontaneous polarization characteristics are a promising alternative for making functional photoferroelectrics. Herein, we report the experimental demonstration of the bulk photovoltaic effect behavior based on the 2D ferroelectric semiconductor α-In$_2$Se$_3$ caused by an out-of-plane polarization induced depolarization field. The α-In$_2$Se$_3$ device exhibits enhanced bulk photovoltaic response in the visible light spectrum owing to its narrow bandgap. It was demonstrated that the generated photovoltaic current density was nearly two orders of magnitude greater than conventional bulk ferroelectric materials. These findings highlight the potential of two-dimensional ferroelectric semiconductor materials for bulk photovoltaic applications in a broad spectral region.

**Introduction**

Photovoltaic energy is regarded as one of the most promising clean energy resources. The pursuit of advanced photovoltaic devices with a high open circuit voltage, high power conversion efficiency, and low-cost fabrication has triggered intense research not only novel materials, but also new photoelectric conversion mechanisms. In contrast to the conversional semiconductor-based photovoltaic effect, which separates photogenerated electron-hole pairs via a PN junction, a bulk material with non-centrosymmetric structure can convert light to electricity that enables the separation of charge carriers via intrinsic polarization electric fileds[4, 5]. This phenomenon is known as Bulk photovoltaic(BPV) effect. [1-3] The BPV effect has numerous advantages, including above bandgap open circuit voltage[6], photovoltaic efficiency cap of Shockley–Queisser limit[7], and relatively simple device structure. However, the BPV study was halted for a long time due to the extremely low power conversion efficiency, which was limited by the wide bandgap (2.7-4 eV) and bulk structure for conventional ferroelectric materials (e.g., $BaTiO_3$, $BiFeO_3$, and so on) [9-11]. Several strategies, such as lowering bandgap[12, 13], reducing ferroelectric layer thickness[14, 15], and optimizing ferroelectric–electrode interfaces [16], have been proposed to improve the efficiency of BPV.

Recently, BPV has received renewed interest with the development of two-dimensional (2D) van der Waals (vdWs) materials. which allow for the combination of low-dimensional nature and exotic physical phenomena, e.g., nonlinear optical effect[17], ferroelectricity[19-21], and flexo-photovoltaic effect[22, 23]. The 2D vdWs BPV response offers appealing characteristics such as high photocurrent density, tunable bandgap, and applicability in low-energy consumption. So far, several 2D materials, such as monolayer MX (M=Ge, Sn, X= S, Se.) [25], monolayer $In_2Se_3$, $CuInP_2S_6$[26], and 2D sliding ferroelectricity, have been described exhibiting BPV effect vis theoretical predictions or experimental evidence. Among them, α-$In_2Se_3$, a 2D ferroelectric semiconductor, has attracted particular interest in photo ferroelectrics[30]. Its unique intercorrelated in-plane and out-of-plane electrical polarization[31-33] and narrow bandgap (around 1.45 eV)[34, 35] enable photovoltaic response that extends to near-infrared[30]. In addition, α-$In_2Se_3$ holds robust ferroelectricity with a Curie temperature of above 500 K[31, 36], making it a candidate for working in high-temperature environment. Many investigations have concentrated on α-$In_2Se_3$. For example, a shift current theory response for monolayer α-$In_2Se_3$ has already been estimated using the first-principle method, and it is higher than that of other 2D materials. However, there are still no experimental examples.

In this article, we experimentally demonstrated the BPV effect in a thin graphite/α-$In_2Se_3$/thin graphite device. The device exhibited an enhanced BPV photocurrent for different laser wavelengths (457 nm, 514 nm and 633 nm) in visible-light range, which is attributed to its lower-dimensional nature and strong out-of-plane polarization intensity. Our findings suggest that 2D ferroelectric semiconductors have the potential to drive the development of advanced optoelectronic devices.

**Results and Discussion**

**Spontaneous polarization for 2H α-In$_2$Se$_3$.**

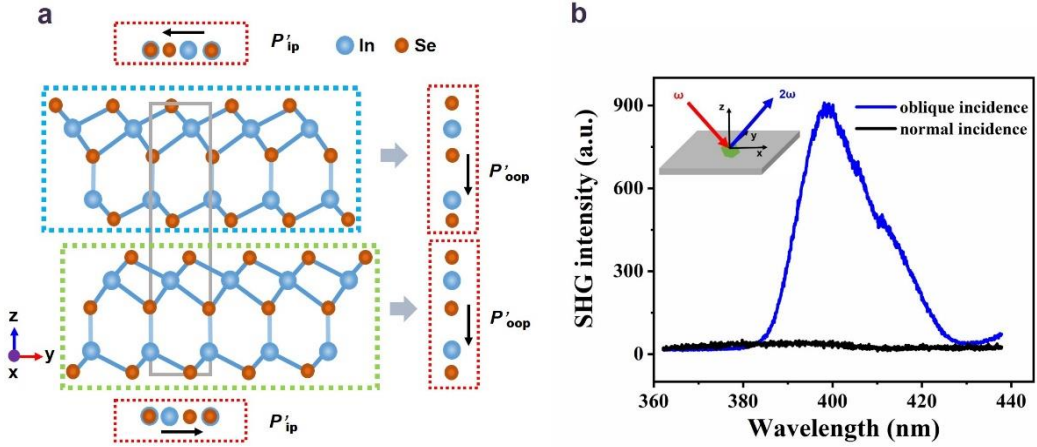

**Figure 1.** Spontaneous polarization for 2H α-In$_2$Se$_3$. a) Side view of bilayer hexagonal α-In$_2$Se$_3$, with blue and orange atoms represented by In and Se balls, respectively. The black arrow in dashed red box represents possible spontaneous polarization. b) SHG signal for In$_2$Se$_3$ flake under normal incidence and oblique incidence, respectively, inset shows the schematic of SHG generation in In$_2$Se$_3$.

The α-In$_2$Se$_3$ crystal possess two possible stacking configurations, known as hexagonal (2H) and rhombohedral (3R) (Figure S1, Supporting Information)[37]. In this work, we study the relatively stable 2H phase, which is non-centrosymmetric[38]. The α-In$_2$Se$_3$ flake were determined by Raman spectroscopy with a splitting peak near 89 cm$^{-1}$, which is interpreted as evidence of 2H stacking[39] (Figure S2, Supporting Information). As shown in **Figure 1**a, each layer in a unit cell for 2H α-In$_2$Se$_3$ has five subatomic layers connected by covalent bonds in the order Se–In–Se–In–Se. The central Se atoms deviate from centrosymmetric position in the covalent bond configuration, leading to in-plane and out-of-plane structural asymmetries. According to the atomic arrangement in a unit cell, net out-of-plane polarization ($P_{oop}$) accumulates as the number of layer increases, resulting from that out-of-plane polarization $P'_{oop}$ for each layer is parallel to z-axis. In contrast, in-plane polarization $P'_{ip}$ in adjacent layers has opposite orientations, leading to the alternating retention of net in-plane polarization ($P_{ip}$) with increasing layer number. Relevant experiments also evidenced that layer-dependent in-plane ferroelectricity in 2H α-In$_2$Se$_3$[40]. Thus, the electrical polarization of 2H α-In$_2$Se$_3$ flake is primarily attributed to out-of-plane polarization. In order to characterize the structural asymmetry, we measured second harmonic generation (SHG) with an excitation laser under oblique and normal incidence, respectively. As shown in Figure 1b, a SHG signal can only be detected under oblique incidence since it contains out-of-plane polarization.

In addition, an electrical transport measurement for a thin graphite/α-In$_2$Se$_3$/thin graphite device is performed. The device's current–voltage $|I_d|$–$V_d$ curve exhibits a hysteretic behavior, suggesting ferroelectricity of 2H α-In$_2$Se$_3$ (Figure S3, Supporting Information).

In order to investigate the BPV response of 2H α-In$_2$Se$_3$, a vertical heterostructure of thin graphite/2H α-In$_2$Se$_3$/thin graphite was fabricated for efficient carrier collection, analogous to conventional ferroelectric photovoltaic devices. As illustrated in **Figure 2**a, α-In$_2$Se$_3$ flake is inserted between two thin graphite(Gr) electrodes, and a designed symmetry contact between the top and down electrodes can exclude short circuit current from Schottky barriers. Spontaneous polarization ***P*** (black arrow) along the z-axis in α-In$_2$Se$_3$ results in polarization (bound) charges at the interfaces (top and bottom). Due to free charges from the graphite electrodes couldn't make up for all of the polarization charges, residual polarization charges will induce a depolarization field ***E*****dp**, which would separate light-generated carriers. As shown in its energy band diagram, the ***E*****dp** induces different vacuum levels at the two interfaces of α-In2Se3 and top/down graphite. Polarization charges at the interfaces alter the chemical potential of top and bottom graphite by doping the same number of free charges. Upon light excitation, electrons and holes are generated, drifted to the interfaces by ***E*****dp**, and finally collected by two graphite electrodes.

Figure 2b displays the current-voltage (*I-V*) characteristic of the α-In$_2$Se$_3$ photovoltaic device at room temperature under laser excitation (red curve) and dark (black curve) conditions. A typical $I_{sc}$ of 0.58 nA is clearly observed upon a laser excitation with the power intensity of 1 μW, whereas the $I_{sc}$ is zero in dark condition (Figure 2b). In general, short circuit current ($I_{sc}$) is a crucial parameter to evaluate the BPV response. Notably, all measurements of $I_{sc}$ and $V_{oc}$ in this work were conducted without any prior poling procedure, thereby the $I_{sc}$ is resulting from spontaneous polarization in α-In$_2$Se$_3$.

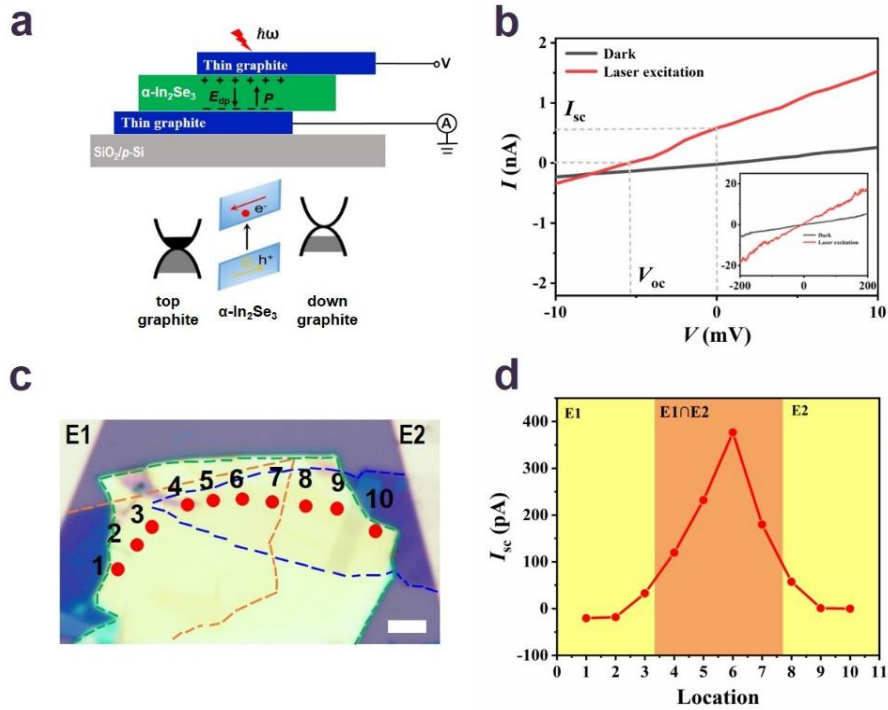

**Figure 2.** BPV effect generation in 2D α-In$_2$Se$_3$ device. a) Schematic of a 2D α-In$_2$Se$_3$ photovoltaic device (D1 device) and the corresponding energy band diagram (down panel). b) I-V characteristics of the D1 device under dark (dark curve) and 1 μW laser excitation (red curve, 633 nm). Inset: a comprehensive view of I-V curves. c) Optical microscope image of the D1 device, comprised of thin graphite on top, 2H α-In$_2$Se$_3$ flake, and thin graphite on down, as denoted by outlined by yellow, green, and blue dashed lines, respectively. Bar of scale, 5 μm. d) $I_{sc}$ measured at various locations of the laser spot labeled on an optical microscope image with a power of 1 μW.

In order to verify the $I_{sc}$ mapping in a 2D α-In$_2$Se$_3$ device, we measured the photocurrent in various regions using a laser spot. The optical image of the 2D α-In$_2$Se$_3$ photovoltaic device is depicted in Figure 2c. The I-V curves at 10 different locations was measured by moving the laser beam from one graphite electrode (E1) to overlapping area (E1∩E2), then to the other graphite electrode (E2). Figure 2d shows the change in $I_{sc}$ extracted from the I-V curves. Due to the hot photocarriers's finite free path length, the photovoltaic reaction $I_{sc}$ appears as the laser spot approaches the overlapping region, grows noticeably stronger within the overlapping region, and fades away as it recedes from the overlapping region. According to the $I_{sc}$ location distribution and the neutralization of the Schottky barrier photovoltaic effect in the overlapping area, we believe that the BPV effect is predominantly responsible for the $I_{sc}$.

**Figure 3**a displays a variation in photocurrent *I* caused by multiple on/off cycles of light illumination. Under light illumination with a power of 20 W and no bias input, a stable photocurrent *I* of 4.1 nA is produced. Moreover, the photocurrent *I* shows minimal temporal variation over multiple cycles from *I–t* curve, suggesting that the photocurrent output of the BPV device is repeatable. Figure 3b presents the *I–V* profiles for laser power densities ranging from 0 to $1.7 \times 10^3$ W cm$^{-2}$. The value of $I_{sc}$ and $V_{oc}$ both grow dramatically with increasing light intensity, the value of $I_{sc}$ can reach up to 16 nA with a laser intensity of $1.7 \times 10^3$ W cm$^{-2}$, which is significantly higher than that of other materials (11 nA@$1.34*10^4$ W cm$^{-2}$ for WS$_2$ nanotubes). Moreover, the $I_{sc}$ linearly depends on light intensity (as depicted in the inset of Figure 3b), satisfying the formula $j=\alpha G_{ijl}P$ for a BPV effect with a linearly polarized light, where *j* is the photocurrent density, *P* is the light intensity, *α* is the absorption coefficient, and $G_{ijl}$ is third-rank piezoelectric tensor. Figure 3c exhibits the $V_{oc}$ and $I_{sc}$ as a function of temperature between 100 and 300 K. $V_{oc}$ and $I_{sc}$, the BPV response, always appear at different temperatures. Specifically, as the device is cooled, the $V_{oc}$ steadily increases and saturates at 150 K, while the $I_{sc}$ remains relatively steady throughout the cooling process. In this case, we only take into account temperature-dependent resistance variation and ignore changes in temperature-dependent absorption coefficients. The total resistance (including bulk resistance $R^*_{bulk}$ and contact resistance $R^*_{contact}$) increases as temperature decreases, according to the *I-V* curves recorded at various temperatures under the dark/light situation (Figure S5, Supporting Information). A simple equivalent circuit model was proposed to explain the dependence of BPV response on temperature, our result aligns fairly well with this model[44].

Figure 3d summarizes the short current density $J_{sc}$ as a function of light intensity $P_{laser}$ at three distinct wavelengths. Overall, the $I_{sc}$ is proportional to *P* in the low power density region (below $2 \times 10^3$ W cm$^{-2}$), and begins to saturate at higher power densities. This is because increasing light intensity can yield more e-h pairs, which causes the $I_{sc}$ to increase, but the limit property of separating e-h pairs slows the increase at high power region. At the same laser power density, the 514 nm laser generated the strongest $I_{sc}$ compared to the 457 nm and the 633 nm lasers. The difference in $I_{sc}$ response corresponds to a distinct absorption coefficient for α-In$_2$Se$_3$ in the visible light region (as displayed in the inset of Figure 3d). The wavelength of 514 nm (2.4 eV) is the closest to the resonant with the In$_2$Se$_3$, and its high absorption intensity produces a larger BPV photocurrent than the wavelengths of 457 nm (2.7 eV) and 633 nm (1.9 eV), which are on the shoulder of the resonant absorption peak. Furthermore, our device exhibited an outstanding BPV response in the visible light range across three wavelengths, whereas the conventional ferroelectric materials responded only in the UV region.

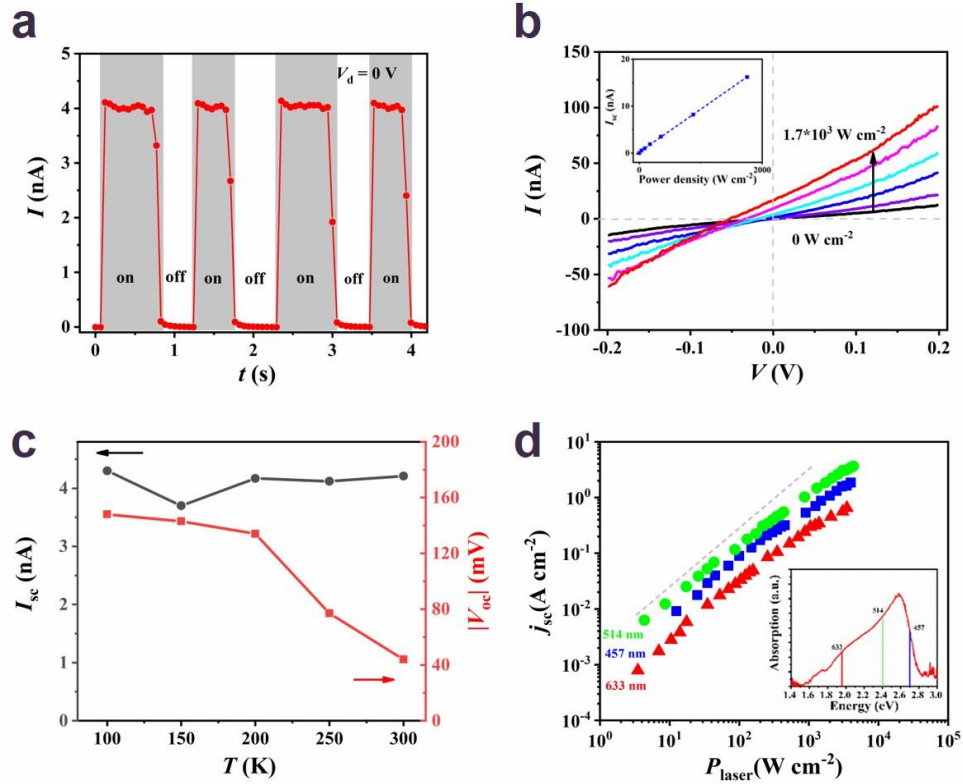

**Figure 3.** The BPV response in α-In$_2$Se$_3$ device (device 2). a) Photocurrent $I$ response recorded by switching on/off of light illumination. b) Temperature dependence of short circuit current (dark circles) and open circuit voltage (red squares). The incident laser in Figures 3a and 3c has the same wavelength and power intensity, 633 nm and 20 μW, respectively. c) $I$–$V$ curves measured under various illumination intensities. In the inset, the $I_{sc}$ as a function of light intensity density is displayed. d) Dependence of the laser power density ($P$) on the short circuit current density ($J_{sc}$) for three distinct lasers with wavelengths of 457 nm (blue squares), 514 nm (green circles), and 633 nm (red triangles). The down right inset illustrates the α-In$_2$Se$_3$ absorption coefficient as a function of photon energy.

**The BPV performance of α-In$_2$Se$_3$**

**Figure 4** presents the short circuit current density ($J_{sc}$) versus the illumination intensity ($P$) for our device and other previously reported materials. The BPV $J_{sc}$ of α-In$_2$Se$_3$ is nearly two orders of magnitude larger than that of conventional ferroelectrical materials (such as, perovskite-type bulk oxides, hybrid oxides, organic molecules, and thin films), while it is smaller than 1D WS$_2$ nanotubes, and comparable to 2D CuInP$_2$S$_6$ ferroelectric in the high power density region. $J_{sc}$ is greatly improved by reducing material thickness and dimensions. In this investigation, the BPV response $J_{sc}$ for 2D ferroelectric semiconductor α-In$_2$Se$_3$ not only bridges the gap between 1D (red shading) and 3D (grey shading) BPV effect materials, but also extends the BPV response to the visible light region compared to the 2D CuInP$_2$S$_6$ (bandgap at 2.9 eV).

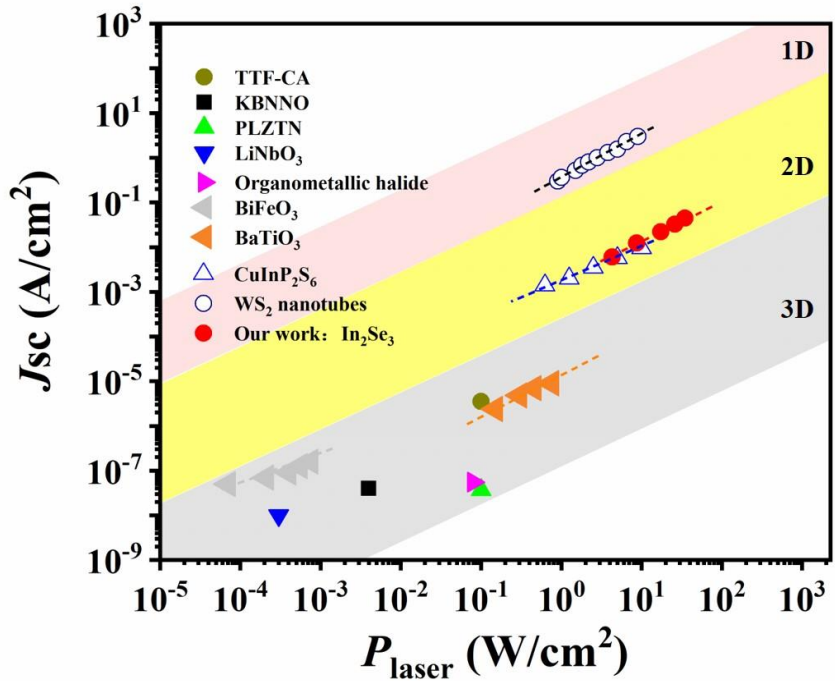

**Figure 4. A summary of the BPV performance for different materials.** Data for other materials are taken from the literature (TTF-CA[44]; KBNNO[13]; PLZTN[46]; LiNbO$_3$[47]; organometallic halide[48]; BiFeO$_3$[10]; BaTiO$_3$[9]; CuInP$_2$S$_6$[26]; and WS$_2$ nanotubes[49]).

**Discussion**

In summary, we report a BPV effect in 2D α-In$_2$Se$_3$ device. The short circuit current density of α-In$_2$Se$_3$ is nearly more than $10^2$ times larger than that conversion ferroelectrical materials due to the inherent out-of-plane electric polarization in the 2D limit. As a result of narrow bandgap for α-In$_2$Se$_3$, it displays an exceptional BPV response in the visible light region. This research makes 2D ferroelectric semiconductors a desirable option for advanced optoelectronics applications, such as solar cells and photodetectors.

## Methods

### Device fabrication

Single crystals of α-In$_2$Se$_3$ and bulk of graphite in this work were provided by HQ graphene and NGS Naturgraphit company, respectively. The α-In$_2$Se$_3$ flakes and thin graphite were successively exfoliated and transferred on a SiO$_2$/*p*-Si substrate. Then, thin graphite/α-In$_2$Se$_3$/thin graphite heterostructures were acquired using an all-dry transfer technique with polydimethylsiloxane(PDMS) stamp in a N$_2$ atmosphere glove box. Metal electrodes were obtained by using laser direct write patterning technique and following electron beam evaporation of Ti/Au (5 nm/30 nm). Afterwards, during the lift-off process, the samples were soaked in NMP at 80°C and maintained for 1 h. Finally, these electrodes were processed with ozone at 80°C for 10 mins.

### AFM, Raman spectrum, and SHG Characterization

Using an atomic force microscope (AFM, Dimension Icon-GB, Bruker), the device's surface morphology was characterized.

Using a Micro-Raman system (Horiba T64000) equipped with a 514 nm excitation laser, Raman spectra were measured. To prevent sample degradation, the laser power is lower than 0.1 mW. In addition, prior to measurement, the system was calibrated using the Raman peak of Si at 520 cm$^{-1}$.

For SHG spectroscopy, the excitation light was from a mode-locked Ti:sapphire laser operating at ~800 nm, 76 MHz repetition rate, 150 fs pulse duration, and 40 μW of average power was used. Using a 50x objective lens and a backscattering configuration, the laser was focused on the surface of sample, and the SHG signal was collected by a cooled CCD spectrometer. For the in-plane polarization measurement, the excitation laser was focused at normal incidence on the surface of the sample. For the out-of-plane polarization analysis, the excitation laser was focused at a tilted angle (approximately 15°-30°) on the surface of sample. The transmissivity of the optical system was carefully calibrated to determine the absolute power level at the focusing plane. The emission spectrum was normalized according to the square of the focused power.

### Electrical Transport and Photovoltaic Measurements

Electrical transport measurement was conducted at room temperature in the vacuum. Keysight source meter (Keithley 4200 Semiconductor parameter analyzer) connected to a cryogenic probe station (CRX-VF, Lake-Shore) was used to measure the I–V characteristics in the dark condition.

Photoelectric measurements were conducted in an optical cryostat (Microscopy Cryostat System $N_2$, Cryo Industries) under high-vacuum condition. These devices were illuminated by three different lasers with wavelengths of a 633 nm (He–Ne laser), 514 nm and 457 nm (two diode laser). The laser spot size is approximately 2.5 μm (633 nm), 1.7 μm (514 nm) and 1.6 μm (457 nm), respectively. The laser was focused on sample using a 50X objective. The applied bias voltage and the output current were controlled via a Keithley 2614B electrometer that was connected to the optical cryostat. The open circuit voltage and short circuit current were derived from the current-voltage curves. The temperature-dependent Id-Vd was carried out in the optical cryostat with heating source.